\documentclass[journal,twoside,onecolumn]{IEEEtran}

\usepackage{cite}
\usepackage{authblk}
\usepackage{setspace}
\usepackage{graphicx}
\usepackage{amssymb}
\usepackage{amsmath}
\usepackage{booktabs}
\usepackage{comment}
\usepackage{caption}
\usepackage{authblk}

\title{Machine Learning-Driven High-Precision Model for $\alpha$-Decay Energy and Half-Life Prediction of Superheavy Nuclei}
\author[1]{Qingning Yuan}
\author[1]{Panpan Qi}
\author[1]{ Xuanpen Xiao}
\author[1]{Xue Wang}
\author[1]{Juan He}
\author[1]{Guimei Long}
\author[1]{Zhengwei Duan}
\author[1]{Yangyan Dai}
\author[1]{Runchao Yan}
\author[1]{Gongming Yu$^{*}$\thanks{Corresponding author: \\ $^{*}$ ygmanan@kmu.edu.cn}}
\author[2]{Haitao Yang$^{\dag}$\thanks{$^{\dag}$yanghaitao205@163.com}}

\affil[1]{College of Physics and Technology, Kunming University, Kunming 650214, China}
\affil[2]{College of Science, Zhaotong University, Zhaotong 657000, China}

\begin{document}

\maketitle

\begin{abstract}
We develop a physics-informed machine-learning framework for predicting $\alpha$-decay energies and half-lives across a broad range of nuclei. The approach is based on an eXtreme Gradient Boosting (XGBoost) regression model and incorporates physically motivated nuclear descriptors constructed from evaluated nuclear data and deformation tables. For half-life prediction, key structure-related features—including magic-number proximity, minimum orbital angular-momentum transfer, isospin asymmetry, and quadrupole deformation—are introduced to represent the dominant mechanisms governing $\alpha$ decay within a data-driven framework. Model performance for both the $Q_{\alpha}$ and half-life prediction tasks is evaluated using five-fold cross-validation, which demonstrates strong predictive accuracy without evident overfitting. Benchmark comparisons with widely used empirical relations, including the Royer formula and the Universal Decay Law (UDL), show that the XGBoost model achieves systematically lower prediction errors while preserving the established systematics of $\alpha$-decay observables. SHapley Additive exPlanations (SHAP) analysis further reveals that the leading contributions from decay-energy information, centrifugal hindrance, and shell-related effects follow physically consistent trends, supporting the interpretability of the learned relationships. These results indicate that gradient-boosting models equipped with physics-guided features provide an accurate and robust framework for describing $\alpha$-decay systematics and for predicting half-lives in regions where experimental data remain scarce.

\end{abstract}

\section{Introduction}
\label{chap:1} 
The study of $\alpha$ decay dates back to the early stages of research on radioactivity. 
In 1899, Rutherford, while investigating uranium-series emissions, first distinguished a radiation component characterized by low penetration but strong ionization and termed it $\alpha$ radiation. Subsequent systematic experiments conducted in 1907–1908 further demonstrated that $\alpha$ particles are doubly charged helium nuclei. This key discovery not only clarified the microscopic nature of radioactive emissions but also established an essential experimental foundation for the development of nuclear-structure models and for understanding nuclear stability. As experimental data continued to accumulate, the regularities underlying $\alpha$ decay gradually emerged, providing a solid basis for systematic investigations of $\alpha$-decay behavior and for the later development of the quantum-tunneling description of nuclear $\alpha$ decay~\cite{rutherford1909xxi,rutherford1899viii,rutherford1908electrical}. Consequently, $\alpha$ decay soon became not only an important subject in early nuclear physics but also a key phenomenon for revealing the structure and stability of atomic nuclei. Building on the accumulating experimental evidence, Geiger and Nuttall (1911) identified a clear correlation between the $\alpha$-decay half-life and the decay energy. This observation led to the formulation of the Geiger–Nuttall law, which states that the logarithm of the $\alpha$-decay half-life exhibits an approximately linear dependence on the inverse square root of the decay energy. The emergence of this empirical relation provided a crucial foundation for subsequent phenomenological descriptions of radioactive decay and significantly advanced the quantitative understanding of nuclear disintegration dynamics~\cite{geiger1911lvii,gamow1928quantentheorie,gurney1928wave}.

Following the empirical establishment of the Geiger--Nuttall law, the subsequent development of quantum mechanics provided a deeper physical interpretation of its underlying regularities. In 1928, Gamow introduced the pioneering quantum-tunneling model, which explained the emission of $\alpha$ particles as a consequence of their ability to penetrate the Coulomb barrier surrounding the daughter nucleus. Within this framework, the decay process is interpreted as the quantum tunneling of a preformed $\alpha$ particle through the potential barrier, naturally reproducing the characteristic exponential dependence of the half-life on the decay energy and providing a microscopic theoretical basis for the Geiger--Nuttall relation. This development marked an important transition toward a quantitative and microscopic understanding of nuclear $\alpha$ decay. Building upon Gamow's theoretical insight, numerous empirical and semi-empirical formulas were subsequently proposed to predict the logarithm of $\alpha$-decay half-lives, among which the Royer formula and the Universal Decay Law (UDL) have been widely applied in nuclear-structure and decay studies~\cite{royer2000alpha,deng2020improved,dong2010alpha,qi2021alpha}. Continuous improvements driven by theoretical advances and the expansion of experimental databases have enhanced the predictive accuracy of these phenomenological models across broad regions of the nuclear chart. Nevertheless, their fixed functional forms limit the extent to which shell effects, nuclear deformation, and angular-momentum hindrance can be fully incorporated, motivating the exploration of approaches that integrate richer physical constraints or data-driven methodologies.Although these empirical formulations incorporate nucleon-parity effects and employ different parameter sets for distinct categories of nuclei, they often exhibit noticeable systematic deviations in low-energy regions or for nuclei far from the valley of stability. This limitation primarily arises because their parameters are fitted using experimental datasets densely concentrated near the stability line and at relatively high $Q_\alpha$ values. In contrast, nuclei in low-$Q_\alpha$ or more exotic regions frequently suffer from larger experimental uncertainties, which reduces both the reliability and the generalizability of these models.Moreover, lower $Q_\alpha$ values correspond to significantly reduced barrier-penetration probabilities that exhibit exponential sensitivity to small variations in $Q_\alpha$. Consequently, even minor deviations in decay energy may lead to orders-of-magnitude discrepancies in half-life predictions. In addition, the simplified treatments of the potential barrier and tunneling dynamics adopted in these empirical approaches, though adequate within the fitted domains, tend to amplify prediction errors as the decay energy decreases, thereby further increasing their systematic deviations~\cite{kelkar2014tunnelin,wang2010fine,liu2024systematic}.

Recent advances in computational technologies have substantially enhanced data processing and analysis capabilities, thereby accelerating the integration of machine-learning techniques across numerous scientific domains. In physics—where the objective is to uncover the fundamental properties of matter and the governing laws of natural phenomena—machine learning has increasingly emerged as a powerful tool for addressing complex problems. Traditional approaches that rely on mathematical modeling, experimental observation, and numerical simulation often face significant challenges when dealing with high-dimensional phase spaces, multiscale systems, or strongly correlated phenomena. Under such circumstances, computational costs become prohibitive and accurate theoretical descriptions may be lacking, ultimately constraining both predictive accuracy and model generalizability~\cite{mumpower2022physically,he2023machine,shinde2018review,lecun2015deep,
zhang2020machine}. In recent years, machine-learning techniques have been increasingly applied to the prediction of $\alpha$-decay energies and half-lives. According to their modeling philosophy and algorithmic structure, existing studies can be broadly categorized into several methodological streams. 

The first category comprises neural-network-based approaches, including artificial neural networks (ANN) and Bayesian neural networks (BNN). These models employ multi-layer nonlinear mappings to capture complex correlations among nuclear-structure descriptors. In particular, BNN frameworks additionally provide uncertainty quantification, which is valuable in extrapolative analyses of nuclear properties~\cite{jin2023bayesian,ma2023simple,saxena2021new,you2022study,liu2025physical,ma2020predictions}.
The second category includes kernel-based methods, such as Support Vector Machines (SVM) and Gaussian Process (GP) regression. These approaches perform nonlinear regression through kernel functions and are widely used in small- and medium-scale regression problems. In nuclear-physics-related applications, kernel-based models have also been explored for modeling complex physical observables. In particular, Gaussian Process models provide a natural probabilistic framework and allow predictive uncertainty estimation, which can be advantageous when dealing with limited experimental data~\cite{jalili2024decay,wang2025exploring,chen2026materials}.
The third category consists of ensemble tree-based methods, including gradient-boosted decision tree algorithms and their variants. These methods iteratively construct decision-tree ensembles to minimize residual errors and have demonstrated strong generalization capability in structured tabular datasets~\cite{li2022deep,liu2026interpretable}.The present work adopts the XGBoost framework, belonging to the ensemble tree-based family, while emphasizing the integration of physics-informed feature design and interpretability analysis. By explicitly incorporating shell-closure proximity, minimum orbital angular-momentum transfer $l_{\min}$, and deformation-related descriptors into the feature set, and by employing SHAP-based attribution analysis, the model not only achieves high predictive accuracy but also establishes a transparent connection between data-driven inference and the underlying physical mechanisms of $\alpha$ decay. Rather than replacing alternative machine-learning paradigms, the objective of this study is to provide a unified framework that balances predictive performance, physical consistency, and interpretability.

Despite recent progress in nuclear-decay modeling, most existing machine-learning approaches still face challenges related to limited precision control and susceptibility to overfitting or underfitting. Furthermore, systematic comparisons with classical empirical formulas across different energy regimes remain insufficient, hindering a comprehensive assessment of model performance. The interpretability of many machine-learning models also requires further improvement, as their internal decision mechanisms often lack clear connections to underlying nuclear-structure physics.

Furthermore, high-quality experimental data for exotic nuclei—particularly those far from the valley of stability or in low decay-energy regions—remain inherently scarce. This scarcity constrains not only the refinement of empirical models but also the development of robust and generalizable data-driven approaches. As a result, constructing models that can deliver accurate and physically consistent predictions under limited-data conditions has become an essential objective in contemporary nuclear-decay research.

To address these challenges, this study introduces an integrated and interpretable machine-learning framework for the simultaneous prediction of $\alpha$-decay energies and half-lives. The model is built upon the XGBoost regression algorithm and incorporates a set of physically motivated nuclear-structure features, including mass number, neutron--proton asymmetry, shell proximity quantified through magic-number distance, and the minimum angular-momentum transfer. Model performance is rigorously benchmarked against traditional empirical formulas, such as the Royer expression and the UDL, using both training and independent test sets. Furthermore, SHAP-based feature attribution is employed to elucidate the dominant physical mechanisms captured by the model, thereby enhancing interpretability and establishing clear connections between data-driven predictions and underlying $\alpha$-decay physics.

This paper is organized as follows. Section~2 describes the XGBoost-based framework employed in this study, including feature construction, data preprocessing, and the overall training procedure. Section~3 presents a detailed analysis of the model's performance in predicting $\alpha$-decay energies and half-lives, together with comparisons against traditional empirical formulas and SHAP-based interpretations of the underlying physical trends. Finally, Section~4 summarizes the principal findings of this work.
    
\section{General Formalism}
\label{chap: Chapter 2} 
This section outlines the systematic modeling workflow adopted in this study, including the fundamental principles of the XGBoost method, the training strategy incorporating an early-stopping mechanism, the feature-engineering procedure, the model parameter configuration and analysis, and the benchmarking methodology against traditional empirical formulas. 
The nuclear properties used in this work—including decay energies, half-lives, and spin–parity assignments—are taken from the evaluated nuclear databases NUBASE2020 and AME2020~\cite{kondev2021nubase2020,huang2021ame,wang2021ame}. 
The quadrupole deformation parameters are adopted from the FRDM2012 deformation tables reported by Möller \textit{et al.}~\cite{moller2016nuclear}. 
All datasets undergo appropriate preprocessing before being utilized for both machine-learning training and empirical-formula evaluation.

Alpha decay is a quantum tunneling process in which a preformed $\alpha$ particle escapes the parent nucleus by penetrating the nuclear potential barrier. Its half-life can be described within the framework of Gamow theory, which employs the Wentzel--Kramers--Brillouin (WKB) approximation to evaluate the tunneling probability. In the standard WKB treatment, the $\alpha$ particle is assumed to be preformed inside the parent nucleus and subsequently tunnels through an effective potential barrier composed of nuclear attraction, Coulomb repulsion, and a centrifugal term. The tunneling probability is predominantly determined by the Gamow factor derived from the WKB approximation~\cite{liu2024systematic,cai2020alpha,hosseini2021systematic}.

\begin{equation}
P = \exp\!\left[
    -\,\frac{2}{\hbar} 
    \int_{R_{\mathrm{in}}}^{R_{\mathrm{out}}}
    \sqrt{2\mu \left( V(r) - Q_{\alpha} \right)}\, dr
\right],
\label{eq:penetration}
\end{equation}
where $\mu$ is the reduced mass of the $\alpha$ particle–daughter nucleus system, 
defined as
\begin{equation}
\mu=\frac{m_{\alpha}m_d}{m_{\alpha}+m_d},
\end{equation}
where $m_{\alpha}$ and $m_d$ denote the masses of the emitted $\alpha$ particle 
and the daughter nucleus, respectively. 
The quantity $V(r)$ represents the total interaction potential between the 
$\alpha$ particle and the daughter nucleus at a relative distance $r$, including 
the Coulomb interaction, the nuclear proximity potential, and the centrifugal 
potential associated with the orbital angular momentum transfer. 
The decay energy $Q_{\alpha}$ denotes the released $\alpha$-decay energy.

The quantities $R_{\mathrm{in}}$ and $R_{\mathrm{out}}$ correspond to the classical 
inner and outer turning points determined from the condition
\begin{equation}
V(r)=Q_{\alpha}.
\end{equation}

Specifically, the inner turning point $R_{\mathrm{in}}$ corresponds to the minimum 
separation between the $\alpha$ particle and the daughter nucleus at the touching 
configuration and is usually approximated as the sum of the nuclear radii of the 
daughter nucleus and the $\alpha$ particle,
\begin{equation}
R_{\mathrm{in}} = R_d + R_{\alpha}.
\end{equation}

The outer turning point $R_{\mathrm{out}}$ is the larger root of the equation 
$V(r)=Q_{\alpha}$ and represents the position where the kinetic energy of the 
emitted particle becomes zero. The interval $[R_{\mathrm{in}}, R_{\mathrm{out}}]$ 
therefore defines the classically forbidden region through which the $\alpha$ 
particle tunnels quantum mechanically.

This formulation explicitly highlights the central role of the decay energy 
$Q_{\alpha}$ in governing the tunneling probability, as variations in 
$Q_{\alpha}$ enter exponentially and therefore exert a dominant influence 
on the resulting half-life. Following Ref.~\cite{liu2024systematic}, the 
$\alpha$-decay half-life can be expressed as:

\begin{equation}
T_{1/2}
= \frac{\ln 2}{\lambda}
= \frac{\ln 2}{\nu S_{c} P},
\label{eq:half_life}
\end{equation}

where $\lambda$ denotes the decay constant, $\nu$ is the assault frequency, 
$P$ is the tunneling probability, and $S_c$ represents the $\alpha$-cluster 
preformation probability. As shown in Eqs.~(\ref{eq:penetration}) and 
(\ref{eq:half_life}), even small variations in the input parameters can lead 
to changes of several orders of magnitude in the calculated half-life. 
This extreme sensitivity arises primarily from the exponential dependence 
of the penetration probability $P$ on the decay energy $Q_{\alpha}$. 
Consequently, any uncertainty in $Q_{\alpha}$ propagates exponentially and 
introduces substantial uncertainty into the predicted half-life~\cite{zhang2008alpha,denisov2009alpha}.

Moreover, because $\alpha$-decay half-lives typically span several orders of magnitude, this broad dynamic range presents significant 
challenges for data modeling and regression analysis. To enhance numerical stability and reduce the influence of extreme values during 
training, the present study adopts the base-10 logarithm of the half-life, $\log_{10} T_{1/2}$, as the regression target~\cite{ahsan2021effect,de2023choice}.
\subsection{\textbf{The Mathematical Fundamentals of eXtreme Gradient Boosting}}
In this work, the Extreme Gradient Boosting (XGBoost) algorithm is employed as the regression framework for learning the nonlinear mapping between nuclear-structure features and decay observables. The selection of XGBoost is motivated by the characteristics of the present problem. The input variables consist of structured, physics-informed nuclear descriptors, including shell proximity, isospin asymmetry, angular-momentum transfer, and quadrupole deformation, for which gradient-boosted decision trees have demonstrated strong performance. 
Moreover, ${\alpha}$-decay datasets are typically moderate in size, and XGBoost provides robust generalization through built-in regularization, shrinkage, and subsampling strategies. Importantly, tree-based ensemble models can naturally capture nonlinear interactions among physical quantities such as $Q_{\alpha}$, nuclear charge number $Z$, and minimum angular momentum $l_{\min}$ without requiring an assumed analytical functional form. XGBoost constructs its predictor as an additive ensemble of regression trees, where the model output for a given sample $\mathbf{x}_i$ is expressed as~\cite{chen2016xgboost,li2023fdpboost}:
\begin{equation}
\hat{y}_i = \sum_{k=1}^{K} f_k(\mathbf{x}_i), \qquad f_k \in \mathcal{F},
\label{eq:additive_model}
\end{equation}
with $\mathcal{F}$ denoting the functional space of CART regression trees. Each tree $f_k$ assigns a constant prediction to samples falling into the same leaf node, and can therefore be written as:
\begin{equation}
f(\mathbf{x}) = w_{q(\mathbf{x})}, 
\qquad q : \mathbb{R}^m \to \{1,2,\dots,L\},
\label{eq:tree_mapping}
\end{equation}
where $L$ is the number of leaf nodes, $q(\mathbf{x})$ specifies the leaf index to which $\mathbf{x}$ is mapped, and $w_j$ is the weight associated with the $j$-th leaf. XGBoost employs a forward stagewise procedure, in which the model at iteration $t$ is obtained by adding a new tree $f_t$ to the previous ensemble, such that~\cite{shree2025alpha}
\begin{equation}
\hat{y}^{(t)}_i = \hat{y}^{(t-1)}_i + f_t(\mathbf{x}_i).
\label{eq:iteration}
\end{equation}

To suppress overfitting and enhance generalization, XGBoost minimizes a regularized objective function composed of a training loss and a structural penalty on the regression trees~\cite{zhang2021study,zuhlke2025tcr}:
\begin{equation}
\mathcal{L}^{(t)}
= \sum_{i} l\!\left(y_i,\hat{y}_i^{(t)}\right)
+ \sum_{k=1}^{t} \Omega(f_k).
\label{eq:objective}
\end{equation}

In the present study the mean-squared-error loss is adopted,
\begin{equation}
l(y_i,\hat{y}_i) = \frac{1}{2}(y_i - \hat{y}_i)^2,
\label{eq:mse}
\end{equation}
while the regularization term penalizes overly complex trees via
\begin{equation}
\Omega(f_t)
= \gamma L_t
+ \frac{1}{2}\lambda \sum_{j=1}^{L_t} w_{tj}^2,
\label{eq:regularization}
\end{equation}
where $L_t$ denotes the number of leaf nodes in tree $f_t$, $\gamma$ controls the creation of new leaves, and $\lambda$ regulates the magnitude of leaf weights. Since only $f_t$ is newly introduced at iteration $t$, the loss contribution of this tree is isolated by expanding the objective using a second-order Taylor approximation around the previous prediction $\hat{y}_i^{(t-1)}$~\cite{zhou2025ensemble,friedman2001greedy}:
\begin{equation}
l\big(y_i,\hat{y}_i^{(t-1)} + f_t(\mathbf{x}_i)\big)
\approx
l\big(y_i,\hat{y}_i^{(t-1)}\big)
+ g_i f_t(\mathbf{x}_i)
+ \frac{1}{2} h_i f_t^2(\mathbf{x}_i),
\label{eq:taylor}
\end{equation}
where
\begin{equation}
g_i = \left.\frac{\partial l}{\partial \hat{y}}\right|_{\hat{y}_i^{(t-1)}}, 
\qquad
h_i = \left.\frac{\partial^2 l}{\partial \hat{y}^2}\right|_{\hat{y}_i^{(t-1)}},
\label{eq:gh}
\end{equation}
and for the MSE loss employed here, $g_i = \hat{y}_i^{(t-1)} - y_i$ and $h_i = 1$. Substituting Eq.~(\ref{eq:taylor}) into Eq.~(\ref{eq:objective}) and discarding constants yields the approximate objective for optimizing the new tree:
\begin{equation}
\tilde{\mathcal{L}}^{(t)}
=
\sum_{i}
\left[
g_i f_t(\mathbf{x}_i)
+ \frac{1}{2} h_i f_t^2(\mathbf{x}_i)
\right]
+ \Omega(f_t).
\label{eq:approx_obj}
\end{equation}

Since $f_t(\mathbf{x}_i)=w_j$ for all samples belonging to leaf $j$, let $I_j=\{\,i \,|\, q(\mathbf{x}_i)=j\,\}$ denote the index set of samples falling into this leaf. Defining the aggregated first- and second-order gradients as
\begin{equation}
G_j = \sum_{i\in I_j} g_i,
\qquad
H_j = \sum_{i\in I_j} h_i,
\label{eq:GH}
\end{equation}
the objective Eq.~(\ref{eq:approx_obj}) becomes
\begin{equation}
\tilde{\mathcal{L}}^{(t)}
=
\sum_{j=1}^{L_t}
\left[
G_j w_j
+
\frac{1}{2}(H_j + \lambda) w_j^2
\right]
+
\gamma L_t.
\label{eq:leaf_loss}
\end{equation}

Minimizing this expression with respect to each leaf weight yields the optimal solution~\cite{sluijterman2025composite,li2022deep}:
\begin{equation}
w_j^{*} = -\frac{G_j}{H_j+\lambda},
\label{eq:w_opt}
\end{equation}
and substituting Eq.~(\ref{eq:w_opt}) into Eq.~(\ref{eq:leaf_loss}) leads to the so-called structure score that evaluates the quality of a given tree structure~\cite{dong2018short}:
\begin{equation}
\tilde{\mathcal{L}}_{\mathrm{struct}}
=
-\frac{1}{2}
\sum_{j=1}^{L_t}
\frac{G_j^{2}}{H_j+\lambda}
+
\gamma L_t.
\label{eq:structure_score}
\end{equation}

When considering a candidate split that divides a leaf $I$ into left and right subsets $I_L$ and $I_R$, with corresponding gradient sums $(G_L,H_L)$ and $(G_R,H_R)$, the improvement in the structure score is quantified by the split gain~\cite{chen2016xgboost}:
\begin{equation}
\mathrm{Gain}
=
\frac{1}{2}
\left[
\frac{G_L^2}{H_L+\lambda}
+
\frac{G_R^2}{H_R+\lambda}
-
\frac{(G_L+G_R)^2}{H_L+H_R+\lambda}
\right]
-
\gamma.
\label{eq:split_gain}
\end{equation}

This quantity serves as the criterion for determining whether a proposed split is beneficial, and therefore governs the tree-growing process in XGBoost.

\subsection{Data input and feature engineering construction}
All nuclear properties employed in this study—including proton number $Z$, neutron number $N$, $\alpha$-decay energy $Q_{\alpha}$, spin--parity assignments $(J^{\pi})$, and experimentally measured half-lives $T_{1/2}$—were taken from the evaluated nuclear databases NUBASE2020 and AME2020~\cite{kondev2021nubase2020,huang2021ame,wang2021ame}. 
Quadrupole deformation parameters $\beta_{2}$ were adopted from the FRDM2012 deformation tables~\cite{moller2016nuclear}. 

Two datasets were constructed for the present study. For the $Q_{\alpha}$ prediction task, a total of 1623 nuclei with experimentally evaluated $Q_{\alpha} > 0$ were retained, covering the region $50 \le Z \le 118$ of the nuclear chart.

\begin{figure}[ht]
    \centering
    \includegraphics[width=0.75\textwidth]{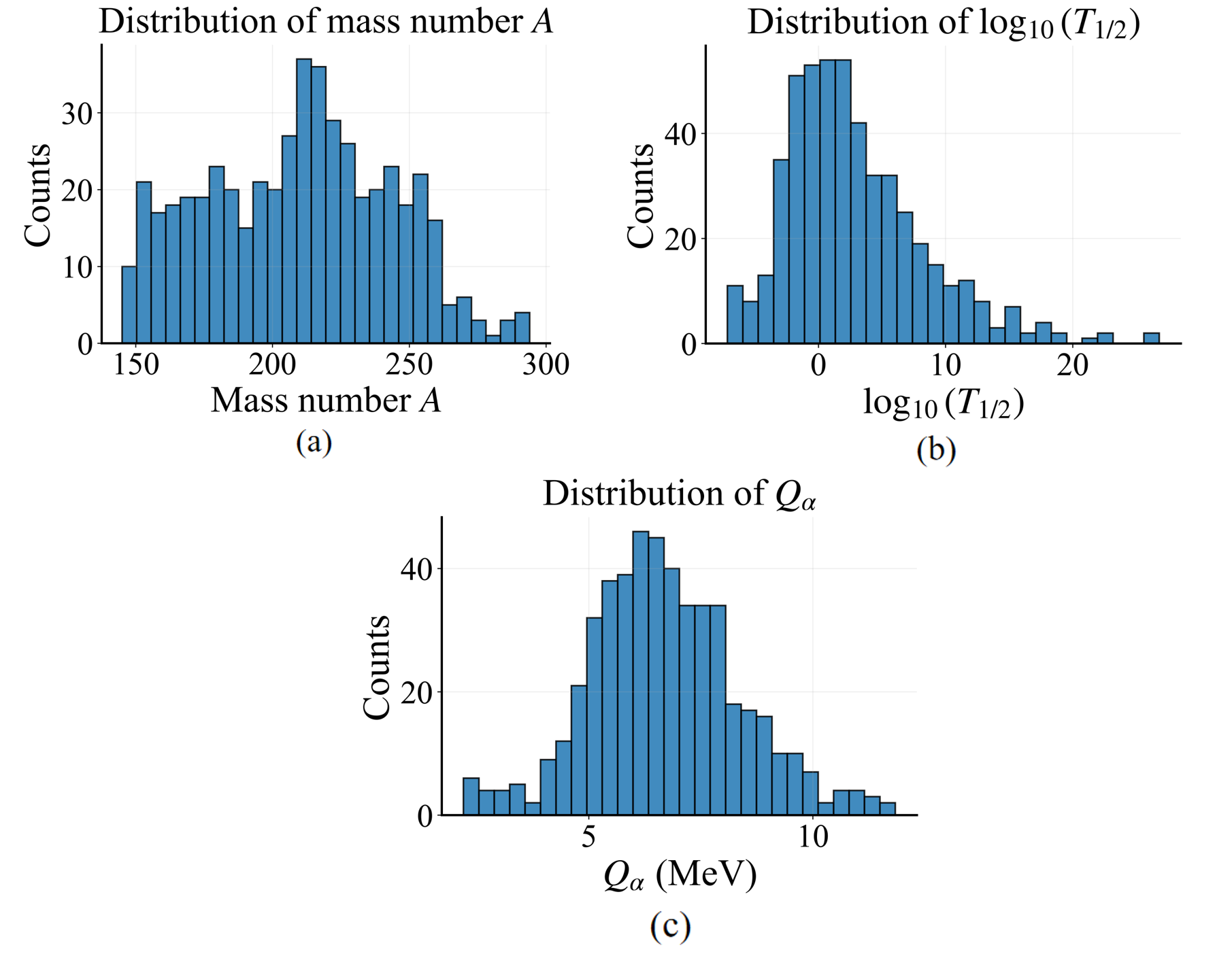}
    \caption{Distributions of (a) mass number $A$, (b) logarithmic half-life $\log_{10}(T_{1/2})$, and (c) $\alpha$-decay energy $Q_{\alpha}$ for the nuclei included in the half-life dataset. The continuous coverage across the considered intervals provides a transparent overview of the data density in different regions of the nuclear chart.}
    \label{fig:counts}
\end{figure}

For the half-life prediction task, a dataset containing 498 nuclei with proton numbers in the range $64 \le Z \le 118$ was constructed. All nuclei included in this dataset possess experimentally measured half-lives $T_{1/2}$ and the complete set of structural quantities required for feature construction. Entries with missing essential nuclear information or undefined structural descriptors were uniformly excluded to ensure consistency of the input feature space.
The corresponding experimental $\alpha$-decay energies span the interval $2.20 \le Q_{\alpha} \le 11.84$ MeV, reflecting the range covered by the available experimental data satisfying the adopted data-quality criteria.
To provide a transparent overview of the data coverage and density across different regions of the nuclear chart, Figure~\ref{fig:counts} presents the distributions of $Q_{\alpha}$, $\log_{10}(T_{1/2})$, and the mass number $A$ for the nuclei included in the half-life dataset.

The overall modeling workflow consists of two conceptually independent components addressing distinct physical observables. 
One component employs an XGBoost-based regression model to predict the $\alpha$-decay energy $Q_{\alpha}$, aiming to capture the nonlinear relationship between nuclear-structure features and decay energy release. The other component independently models the $\alpha$-decay half-life by incorporating experimentally measured decay energies together with nuclear-structure parameters, enabling a direct investigation of the physical factors governing $T_{1/2}$ without introducing uncertainty propagation from decay-energy prediction.

\subsubsection{Decay Energy Prediction}

To predict the decay energy, five nuclear-structure descriptors were employed as input variables for the model. Their definitions are summarized in Table~\ref{tab:decay_energy_features}.

\begin{table}
\centering
\caption{Decay-energy prediction related features used in the model.}
\renewcommand{\arraystretch}{1.4}
\begin{tabular}{cc}
\hline
\hline
Name & Relevant Calculations \\
\hline
Mass number & $A$ \\
Proton number & $Z$ \\
Neutron number & $N$ \\
Proton--neutron ratio & $Z/N$ \\
Relative neutron excess & $(N-Z)/A$ \\
\hline
\hline
\end{tabular}
\label{tab:decay_energy_features}
\end{table}

These variables reflect the fundamental nuclear properties and nucleon configurations within atomic nuclei. Although the mass number $A$ is mathematically related to the proton number $Z$ and the neutron number $N$ through $A = Z + N$, these quantities are not redundant from a physical perspective. Instead, they correspond to different aspects of nuclear structure: the proton number $Z$ is closely associated with Coulomb effects and proton shell structure, the neutron number $N$ characterizes isotopic systematics and neutron shell properties, while the mass number $A$ is related to the overall nuclear size and mass scale. As key structural variables governing nuclear binding and decay behavior, they exhibit pronounced nonlinear correlations with the decay energy, a relationship well supported by both the liquid-drop model and shell-model frameworks~\cite{royer2008coefficients,zhang2006alpha}. Additional tests indicate that removing any one of these variables, while keeping the remaining features unchanged, leads to a slight decrease in predictive accuracy. This result suggests that retaining these fundamental nuclear descriptors helps the model capture systematic nuclear-structure trends more effectively. Maintaining a consistent set of input features across light, heavy, and superheavy mass regions further improves the generalizability of the model over the entire nuclear chart.

In this study, nuclides with positive decay energies ($Q_{\alpha} > 0$) were collected to construct the dataset used for the $Q_{\alpha}$ prediction task. 
To obtain a reliable estimate of the predictive performance, a five-fold cross-validation strategy was adopted. 
Specifically, the dataset was randomly divided into five mutually exclusive subsets of approximately equal size. 
In each fold, four subsets were used for model training and the remaining subset was used for testing, so that every nucleus was evaluated once as independent test data. 
The final performance metrics were reported as the mean and standard deviation over the five folds. 
The results indicate that the proposed model possesses strong predictive capability, and the detailed numerical results together with the SHAP-based interpretability analysis will be discussed in the following section.

\subsubsection{Half-life Prediction}

This stage constitutes the core component of the methodological framework developed in this study. 
In this part, the $\alpha$-decay half-life regression model is constructed by incorporating the decay energy $Q_{\alpha}$ together with multiple nuclear-structure features. 
The $Q_{\alpha}$ values are taken from the AME2020 atomic mass evaluation, and the half-life data are adopted from the NUBASE 2020 database, both of which are based on experimentally measured or systematically evaluated nuclear data. 
The machine-learning model directly performs regression on the experimentally measured $\log_{10} T_{1/2}$ values, without employing residual corrections to empirical formulas or theoretical model predictions as training targets. 
The model takes physics-informed nuclear descriptors as input features and establishes a nonlinear mapping to the decay observables. 
The deformation parameters are extracted from the FRDM(2012) nuclear mass and deformation database, and are used solely as structural input features rather than for generating decay observables. 
All feature parameters employed in the half-life prediction task are summarized in Table~\ref{tab:half_life_features}.

\begin{table}
\centering
\caption{Physics-informed feature set used in the half-life prediction model.}
\renewcommand{\arraystretch}{1.4}
\begin{tabular}{cc}
\hline
\hline
Name & Relevant Calculations \\
\hline

Coulomb--energy coupling descriptor & $Z_1/\sqrt{Q_{\alpha}}$ \\
Proton number of the parent nucleus & $Z_1$ \\
Neutron number of the parent nucleus & $N_1$ \\
Mass number of the parent nucleus & $A_1$ \\
Neutron--proton ratio of the parent nucleus & $N_1/Z_1$ \\
Relative neutron excess of the parent nucleus & $(N_1-Z_1)/A_1$ \\
Minimum orbital angular momentum & $\ell_{\min}$ \\
Quadrupole deformation term & $\sqrt{\kappa_2\beta_2}$ \\
Distance to the nearest proton shell closure & $\lvert Z_1-\text{magic}\rvert_{\min}$ \\

\hline
\hline
\end{tabular}
\label{tab:half_life_features}
\end{table}

In the feature system constructed for this study, the mass number $A_1$, neutron number $N_1$, and proton number $Z_1$ of the parent nucleus serve as fundamental structural quantities. These variables determine the position of a nucleus on the nuclear chart and encode essential information on nuclear size, isospin asymmetry, and shell structure, thereby playing an important role in the description of $\alpha$-decay dynamics. 

Beyond these basic attributes, several additional physically motivated quantities are introduced to characterize the key mechanisms governing the decay process. These include the Coulomb–energy coupling term $Z_1/\sqrt{Q_\alpha}$, the minimum orbital angular momentum $\ell_{\min}$ required for $\alpha$ emission, indicators of isospin asymmetry, shell-structure descriptors, and deformation-related parameters. Together, these quantities allow the model to incorporate both decay-energy dependence and nuclear-structure effects within a unified feature framework~\cite{seif2019additional,sheline1991alpha,delion2007alpha}.

Among these features, the minimum orbital angular momentum $\ell_{\min}$ required for $\alpha$ emission represents an important descriptor of the centrifugal barrier. Its value is determined by the coupling between the spins ($j_p, j_d$) and parities ($\pi_p, \pi_d$) of the parent and daughter nuclei and follows the selection rules~\cite{saxena2021modified}:

\begin{equation}
\ell_{\min}=
\begin{cases}
\Delta_j,   & \text{if }\Delta_j \text{ is even and } \pi_p=\pi_d,\\[4pt]
\Delta_j+1, & \text{if }\Delta_j \text{ is even and } \pi_p\neq \pi_d,\\[4pt]
\Delta_j,   & \text{if }\Delta_j \text{ is odd and } \pi_p\neq \pi_d,\\[4pt]
\Delta_j+1, & \text{if }\Delta_j \text{ is odd and } \pi_p=\pi_d,
\end{cases}
\qquad
\Delta_j = |j_p - j_d|.
\label{eq:lmin_rules}
\end{equation}

This quantity represents the lowest orbital angular momentum transfer compatible with angular-momentum and parity conservation. Larger values of $\ell_{\min}$ increase the effective barrier through the centrifugal term, thereby reducing the tunneling probability and extending the $\alpha$-decay half-life. The inclusion of $\ell_{\min}$ therefore enables the model to account for the well-established angular-momentum hindrance effect observed in $\alpha$-decay systematics.

The relative neutron excess $(N_1 - Z_1)/A_1$ and the neutron-to-proton ratio $N_1/Z_1$ are introduced to characterize the isospin asymmetry of the parent nucleus. These quantities are closely related to the symmetry-energy term in the liquid-drop model and therefore provide important information on nuclear binding and structural stability. Their variations reflect the degree to which a nucleus deviates from the valley of stability. 

For neutron-rich nuclides located far from stability, isospin asymmetry significantly influences the shape of the decay barrier and the preformation probability of the $\alpha$ cluster, thereby affecting the decay behavior~\cite{cui2022improved,otsuka2020evolution}. Incorporating these asymmetry-related descriptors into the feature set allows the model to better capture systematic trends along isotopic chains and improves the prediction accuracy for exotic and weakly bound nuclei.

To characterize the proximity of the parent nucleus to proton shell closures, we introduce the minimum proton magic-number distance $\Delta Z_{\text{magic}}$, which quantifies the influence of shell structure on the $\alpha$-decay half-life~\cite{nakada2019irregularities,sorlin2008nuclear,amiel1975odd}. The proton magic numbers considered in this work are 2, 8, 20, 28, 50, 82, and 114. Nuclei located near shell closures generally possess enhanced binding energy and increased structural stability, which significantly affect the shape of the decay barrier and the preformation probability of the $\alpha$ cluster. Thus, this feature effectively strengthens the model’s sensitivity to shell effects.

To further examine the impact of nuclear deformation on the $\alpha$-decay 
half-life, a deformation-related correction term $\sqrt{\kappa_2 \beta_2}$ 
is introduced. Not all nuclei possess perfectly spherical shapes; for the 
majority of medium and heavy nuclei, spontaneous symmetry breaking leads 
to noticeable deformation, which manifests as a pronounced electric 
quadrupole moment. The nuclear surface is often parametrized by standard 
deformation variables, among which the quadrupole deformation parameter 
$\beta_2$ is the simplest and most influential, as it governs the dominant 
deformation effects on the decay barrier \cite{stone2016table,ma2022high,
butler2023studies,garrett2022experimental,qian2012shape,manhas2005proximity}. 
In the present work, $\beta_2$ refers to the quadrupole deformation 
parameter of the parent nucleus.

To describe the contribution of quadrupole deformation in the parent nucleus, different values of the coefficient $\kappa_2$ are assigned according to the deformation type: $\kappa_2 = 2$ for prolate shapes ($\beta_2 > 0$), $\kappa_2 = -1$ for oblate shapes ($\beta_2 < 0$), and $\kappa_2 = 0$ when the nucleus is nearly spherical ($\beta_2 = 0$). This assignment is designed to embed, within the empirical modeling framework, the qualitative influence of distinct deformation modes on the $\alpha$-decay half-life.

By combining the deformation sign information ($\kappa_2$) with the deformation magnitude ($\beta_2$), the constructed term $\sqrt{\kappa_2 \beta_2}$ effectively captures the modification of the half-life arising from deformation-induced changes in the Coulomb barrier \cite{denisov2025alpha,denisov2024empirical,yazdankish2023improved}. Incorporating this feature enhances the model’s sensitivity to barrier-geometry effects and improves its physical consistency across nuclei exhibiting different deformation characteristics.

The selection of these features is grounded in well-established nuclear-physics models, enabling a deeper understanding of nuclear structural stability and the underlying mechanisms governing $\alpha$-decay. By systematically incorporating key physical effects—such as shell structure, isospin asymmetry, angular-momentum conservation, and nuclear deformation—into the feature space, the model achieves enhanced predictive accuracy and improved physical consistency. This physically informed feature design further strengthens the generalization capability of our half-life framework and provides a more robust basis for interpreting the model’s predictions.

To ensure a robust and reliable evaluation of model performance, the same 
five-fold cross-validation strategy described in the $Q_{\alpha}$ prediction 
task was adopted for the half-life regression model. In each fold, four subsets 
were used for training and the remaining subset was used for testing, and the 
final performance metrics were reported as the mean and standard deviation 
over the five folds.
Within each cross-validation fold, 10\% of the training data were further 
separated as a validation subset. Early stopping was implemented based on the 
validation RMSE, and the training process was terminated if no improvement 
was observed for 200 consecutive boosting rounds. The model parameters 
corresponding to the lowest validation RMSE were retained. In addition, a fixed 
random seed was used throughout the training process to ensure the 
reproducibility of the results. This procedure helps prevent overfitting and 
improves the stability of the training process.
The detailed predictive performance of the model, together with the SHAP-based 
interpretability analysis, is presented in Section~3.

During model training, the hyperparameters of the XGBoost model were kept fixed and consistently applied across all cross-validation folds to ensure the stability and reproducibility of the training procedure. The model was trained using the Pseudo-Huber loss function together with a histogram-based tree construction algorithm.
The main hyperparameters include the learning rate, maximum tree depth, subsample ratio, column sampling ratio, minimum child weight, and regularization coefficients. In addition, early stopping based on the validation RMSE was employed to prevent overfitting during training. The complete hyperparameter configuration is summarized in Table~\ref{tab:xgb_params}, and the same settings were used across all cross-validation folds.

\begin{table}
\centering
\caption{Hyperparameter configuration of the XGBoost model used in the 
$\alpha$-decay half-life prediction task.}
\renewcommand{\arraystretch}{1.4}
\begin{tabular}{cc}
\hline
\hline
Hyperparameter & Value \\
\hline
Number of estimators (n\_estimators) & 1600 \\
Learning rate & 0.03 \\
Maximum tree depth (max\_depth) & 4 \\
Minimum child weight (min\_child\_weight) & 6 \\
Subsample ratio (subsample) & 0.85 \\
Column sample by tree (colsample\_bytree) & 0.85 \\
Gamma ($\gamma$) & 0.10 \\
L1 regularization ($\alpha$) & 0.8 \\
L2 regularization ($\lambda$) & 5.0 \\
Early stopping rounds & 200 \\
\hline
\hline
\end{tabular}
\label{tab:xgb_params}
\end{table}

\subsection{Empirical Formulas (Royer and Universal Decay Law)}

In studies of $\alpha$-decay half-lives, empirical formulas with compact analytical forms are widely employed for rapid estimation owing to their computational efficiency. Such models frequently serve as benchmark references for theoretical frameworks, where nuclide classification schemes are often incorporated to improve predictive performance. Although they do not explicitly describe all microscopic mechanisms, they continue to play a central role in systematics-based investigations of nuclear decay.

In this work, we adopt two of the most widely used and continuously refined empirical formulations: the Royer formula and the Universal Decay Law (UDL). A brief introduction to their analytical structures is provided in this section, and both will be utilized as comparative benchmarks in subsequent analyses to evaluate the physical consistency and predictive capability of our machine-learning framework.

\subsubsection{Royer Formula}

The Royer formula, originally proposed by G.~Royer, is derived from the liquid-drop model and subsequently fitted to extensive experimental datasets \cite{royer2000alpha,deng2020improved,dong2010alpha,moller2016nuclear}. This empirical expression is applicable to nuclides with different proton--neutron parity combinations, and its general form is given by:

\begin{equation}
\log_{10} (T_{1/2}^{\text{Royer}}(s))
  = a + b A^{1/6} \sqrt{Z}
    + \frac{c Z}{\sqrt{Q_\alpha}},
\label{eq:Royer}
\end{equation}
to extend the applicability of the Royer expression to $\alpha$-decay processes involving non-zero angular momentum transfer, G.~Royer and collaborators introduced an angular-momentum correction term. Furthermore, to account for the influence of nuclear deformation on the $\alpha$-decay half-life, an additional quadrupole deformation term associated with the parent nucleus is incorporated into the formulation. Accordingly, in the present study we adopt the modified Royer expression reported in Refs.~\cite{wang2010fine,moller2016nuclear,qian2012shape,manhas2005proximity,yazdankish2023improved} as the empirical benchmark for comparison.

\begin{equation}
\log_{10} (T_{1/2}^{\text{Royer}}(s))
= a + b A^{1/6}\sqrt{Z}
+ \frac{c Z}{\sqrt{Q_\alpha}}
+ \frac{\ell(\ell+1)}{\sqrt{(A-4)(Z-2)}\,A^{2/3}}
+ d \sqrt{\kappa_2 \beta_2}\,\frac{Z}{\sqrt{Q_\alpha}} ,
\label{eq:modified_Royer}
\end{equation}
Here, $A$ denotes the mass number of the parent nucleus, $Z$ is the proton number, and $Q_\alpha$ represents the decay energy. 
The determination of $\kappa_{2}$ follows exactly the same convention described earlier, and the same rule also applies to the 
UDL formulation. The coefficients $a$, $b$, $c$, and $d$ are fitted parameters whose values depend on the even--odd structure 
of the parent nucleus. The corresponding parameter sets for the four nucleon-parity classifications (e--e, e--o, o--e, and o--o) 
are summarized in Table~\ref{tab:Royer_coeff}.

\begin{table}
\centering
\caption{Coefficients in the Royer formula.}
\renewcommand{\arraystretch}{1.4}
\begin{tabular}{ccccc}
\hline
\hline
Nuclei & $a$ & $b$ & $c$ & $d$ \\
\hline
e--e & -25.3100 & -1.1629 & 1.5864 & -0.0106 \\
e--o & -26.6500 & -1.0859 & 1.5848 & -0.0186 \\
o--e & -25.6800 & -1.1423 & 1.5920 & 0.0156 \\
o--o & -20.4800 & -1.1130 & 1.6971 & -0.0223 \\
\hline
\hline
\end{tabular}
\label{tab:Royer_coeff}
\end{table}

\subsubsection{Universal Decay Law Formula}

The UDL formula offers several significant advantages. Unlike purely empirical expressions, it is derived from a clear physical foundation that encapsulates the essential mechanism of barrier penetration in radioactive decay. Moreover, the UDL provides a unified description for both $\alpha$ decay and heavier cluster radioactivities, enabling consistent application across different decay modes within a single theoretical framework. It also exhibits strong predictive performance for even--even, odd--even, and odd--odd nuclei, and can be reliably extrapolated to the superheavy region, offering valuable insights into the structural stability of newly synthesized elements~\cite{qi2021alpha,yazdankish2023improved,zhao2022unified}.

\begin{equation}
\log_{10} (T_{1/2}^{\mathrm{UDL}}(s))  =
a \frac{\sqrt{\mu}\, Z_\alpha Z_d}{\sqrt{Q_\alpha}}
+ b \left[ \sqrt{\mu}\, Z_\alpha Z_d \left( A_\alpha^{1/3} + A_d^{1/3} \right) \right]^{1/2}
+ c + d\, \ell(\ell+1) + e I + f I^{2}.
\end{equation}

Similarly, by incorporating the quadrupole deformation parameter $\beta_{2}$ into the original UDL formulation, nuclear shape effects 
can be further embedded within the empirical description. This modification enables a more comprehensive representation of deformation-induced changes in the decay barrier and consequently the half-life. The resulting expression takes the following form \cite{moller2016nuclear,manhas2005proximity,qian2012shape,yazdankish2023improved}:

\begin{equation}
\begin{aligned}
\log_{10} (T_{1/2}^{\mathrm{UDL}}(s)) 
={}&\, a\sqrt{\mu Z_{\alpha} Z_{d}}\,\frac{1}{\sqrt{Q_{\alpha}}}
   + b\left[ \sqrt{\mu Z_{\alpha} Z_{d}} \left( A_{\alpha}^{1/3} + A_{d}^{1/3} \right) \right]^{1/2} \\
 &\quad + c + d\,\ell(\ell+1) + eI + fI^{2} 
   + g\sqrt{\kappa_{2}\beta_{2}}\,\frac{Z}{\sqrt{Q_{\alpha}}}\, .
\end{aligned}
\label{eq:UDL_deform}
\end{equation}

Among the terms appearing in the UDL expression, $Z_{\alpha}=2$ and 
$A_{\alpha}=4$ denote the proton number and mass number of the emitted 
$\alpha$ particle, respectively. The quantities $Z_d$ and $A_d$ represent 
the proton number and mass number of the daughter nucleus, defined as 
$Z_d = Z - Z_{\alpha}$ and $A_d = A - A_{\alpha}$. 
The quantity $I = (N - Z)/A$ characterizes the relative neutron excess of 
the parent nucleus and reflects the isospin asymmetry of the nuclear 
system. The fitting coefficients of the UDL formula depend on the 
even--odd parity combination of the parent nucleus's proton and neutron 
numbers, corresponding to four distinct cases summarized in 
Table~\ref{tab:udl_coefficients}.

Both empirical formulations---the Royer model and the UDL---have undergone extensive validation in $\alpha$-decay systematics. Although they cannot fully reproduce all microscopic mechanisms, they provide essential phenomenological frameworks for understanding global decay trends. Building upon these foundations, we subsequently generate half-life predictions using both empirical models and perform a comprehensive comparative analysis against the results obtained from our machine learning framework.

\begin{table}
\centering
\caption{Coefficients used in the empirical $\alpha$-decay half-life formula for different odd--even combinations of parent nuclei.}
\renewcommand{\arraystretch}{1.4}
\begin{tabular}{cccccccc}
\hline
\hline
Nuclei & $a$ & $b$ & $c$ & $d$ & $e$ & $f$ & $g$ \\
\hline
e--e & 0.4226 & -0.5582 & -25.116 & 0 & 2.5165 & -27.929 & -0.0183 \\
e--o & 0.4290 & -0.5504 & -25.916 & 0.04578 & -15.408 & 43.457 & -0.0218 \\
o--e & 0.4275 & -0.5428 & -24.855 & 0.0135 & -16.031 & 32.542 & -0.0085 \\
o--o & 0.4341 & -0.5424 & -26.0589 & 0.0401 & -6.7131 & 14.46 & -0.0235 \\
\hline
\hline
\end{tabular}
\label{tab:udl_coefficients}
\end{table}

\section{Results and Discussion\label{Results and Discussion}}
\subsection{Decay Energy Prediction Performance}

To quantitatively evaluate the predictive performance of the decay-energy model, 
the key error metrics obtained from the five-fold cross-validation procedure 
are summarized in Table~\ref{tab:Q_metrics}. The results are reported as the 
mean and standard deviation over the five folds. The model achieves consistently 
high predictive accuracy with only small differences between the training and 
testing folds, indicating strong generalization capability and no evidence of 
significant overfitting.

\begin{table}
\centering
\caption{Five-fold cross-validation results of the XGBoost model for $Q_{\alpha}$ prediction. 
The error metrics (RMSE and MAE) are reported in MeV.}
\renewcommand{\arraystretch}{1.4}
\begin{tabular}{ccc}
\hline
\hline
Metric & Training (mean $\pm$ std) & Testing (mean $\pm$ std) \\
\hline
RMSE & $0.2403 \pm 0.0121$ & $0.2971 \pm 0.0430$ \\
MAE  & $0.1147 \pm 0.0053$ & $0.1765 \pm 0.0114$ \\
$R^{2}$ & $0.9936 \pm 0.0006$ & $0.9901 \pm 0.0024$ \\
\hline
\hline
\end{tabular}
\label{tab:Q_metrics}
\end{table}

The results show that the XGBoost model maintains very high predictive 
accuracy across different data partitions. Although the testing errors are 
slightly larger than the training errors, the overall error level remains low 
and the standard deviations across the five folds are small, demonstrating the 
stability and robustness of the model when applied to unseen nuclei.

\begin{figure}[ht]
\centering
\includegraphics[width=0.70\textwidth]{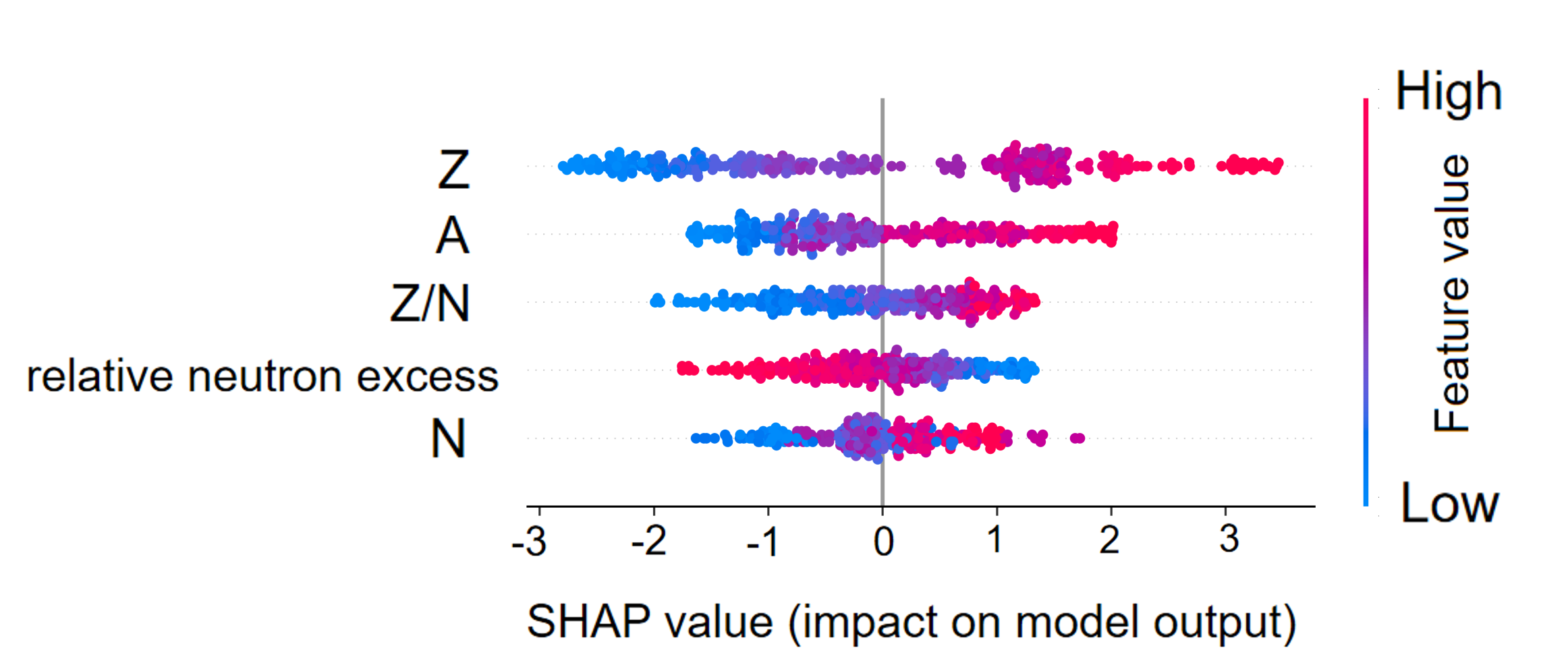}
\caption{SHAP summary plot showing the feature-importance distribution in the 
decay-energy prediction model. The horizontal axis represents SHAP values, 
indicating the contribution of each feature to the model output. Each point 
corresponds to a single prediction sample and is colored according to the 
feature value from low (blue) to high (red).}
\label{fig:eng_shap}
\end{figure}

Having obtained stable and physically consistent decay-energy predictions, it 
is important to further investigate the physical factors that the model relies 
on during the learning process. To this end, the SHAP framework was employed to 
perform a feature-importance analysis, enabling a quantitative evaluation of 
the contributions of different nuclear-structure descriptors to the predicted 
decay energies. The interpretability results are shown in 
Figure~\ref{fig:eng_shap}.

Figure~\ref{fig:eng_shap} presents the SHAP feature-importance distribution for 
the decay-energy prediction task. The model primarily relies on the proton 
number $Z$, neutron number $N$, mass number $A$, and symmetry-energy-related 
quantities such as $Z/N$ and the relative neutron excess. The importance 
ranking of these features is consistent with expectations from the liquid-drop 
model and shell-model descriptions of nuclear binding energies, indicating 
that the model successfully captures the known systematic behavior governing 
$Q_{\alpha}$ values.

Overall, the essential nuclear-structure information relevant to $Q_{\alpha}$ 
is effectively captured, demonstrating the model’s capability to learn 
physically meaningful correlations within decay-energy systematics.

\subsection{Half-Life Prediction Performance}

To quantify the baseline performance of traditional analytical approaches, 
we evaluate the Royer and UDL formulas on the full $\alpha$-decay dataset 
considered in this work. The corresponding global prediction errors are 
summarized in Table~\ref{tab:royer_udl_perf}. These values represent the 
typical level of accuracy achievable by widely used empirical relations. 
As expected, the UDL formula exhibits substantially better overall 
performance than the Royer expression, reflecting its more comprehensive 
incorporation of decay-systematics trends. These baseline results provide 
a quantitative reference for evaluating the performance of the 
machine-learning model introduced below.

\begin{table}
\centering
\caption{Overall prediction errors of the Royer and UDL formulas on the full 
$\alpha$-decay dataset, evaluated for $\log_{10}(T_{1/2}(s))$.}
\renewcommand{\arraystretch}{1.4}
\begin{tabular}{ccc}
\hline
\hline
Model & RMSE & MAE \\
\hline
UDL formula & 0.8887 & 0.6839 \\
Royer formula & 3.0897 & 1.4727 \\
\hline
\hline
\end{tabular}
\label{tab:royer_udl_perf}
\end{table}

For the half-life prediction task, a regression model is constructed by combining decay-energy information with multiple nuclear-structure 
features. To systematically evaluate the predictive performance of the model, three key metrics—$R^{2}$, RMSE, and MAE—are calculated within a five-fold cross-validation framework. The results, summarized in Table~\ref{tab:hl_perf}, show that the model maintains consistently high 
predictive accuracy across different data partitions. The relatively small variations of the evaluation metrics among the folds indicate stable model performance under different data splits. Meanwhile, the model also retains high predictive accuracy for nuclei that are not involved in the training process, suggesting good generalization capability of the proposed approach.

\begin{table}
\centering
\caption{Five-fold cross-validation results of the XGBoost model for 
$\alpha$-decay half-life prediction. The error metrics (RMSE and MAE) are 
evaluated for $\log_{10}(T_{1/2}(s))$.}
\renewcommand{\arraystretch}{1.4}
\begin{tabular}{ccc}
\hline
\hline
Metric & Training (mean $\pm$ std) & Testing (mean $\pm$ std) \\
\hline
RMSE & $0.5231 \pm 0.0630$ & $0.7845 \pm 0.1622$ \\
MAE  & $0.3648 \pm 0.0112$ & $0.4882 \pm 0.0469$ \\
$R^{2}$ & $0.9906 \pm 0.0023$ & $0.9780 \pm 0.0137$ \\
\hline
\hline
\end{tabular}
\label{tab:hl_perf}
\end{table}

\begin{figure}[ht]
\centering
\includegraphics[width=0.75\textwidth]{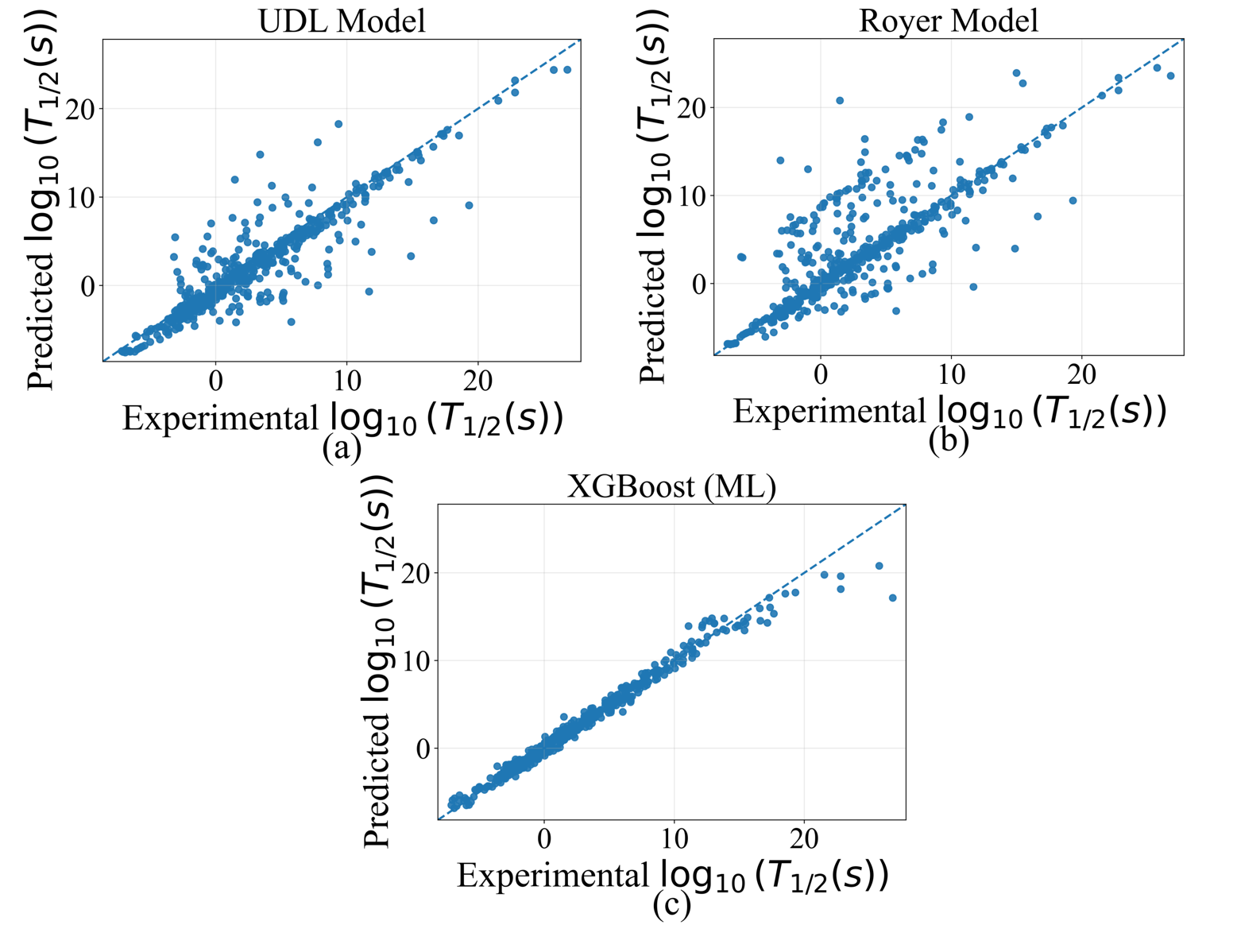}
\caption{
Comparison between predicted and experimental values of $\log_{10}(T_{1/2}(s))$ obtained using (a) the UDL empirical formula, (b) the Royer empirical formula, and (c) the XGBoost model. The horizontal axis represents the experimental values and the vertical axis denotes the predicted values. For the XGBoost model, the points correspond to test predictions obtained from five-fold cross-validation. The dashed line indicates the ideal relation $y=x$.
}
\label{fig:total}
\end{figure}

The improvement of the XGBoost model relative to the Royer and UDL formulas can be clearly observed in both the quantitative performance metrics (Table~\ref{tab:hl_perf}) and the model–data comparison plots shown in Figure~\ref{fig:total}. In particular, the machine-learning predictions exhibit a noticeably tighter distribution around the ideal relation $y=x$, indicating that the model more accurately captures the systematic behavior of $\alpha$-decay half-lives across the considered nuclear region.

\begin{figure}[ht]
\centering
\includegraphics[width=0.75\textwidth]{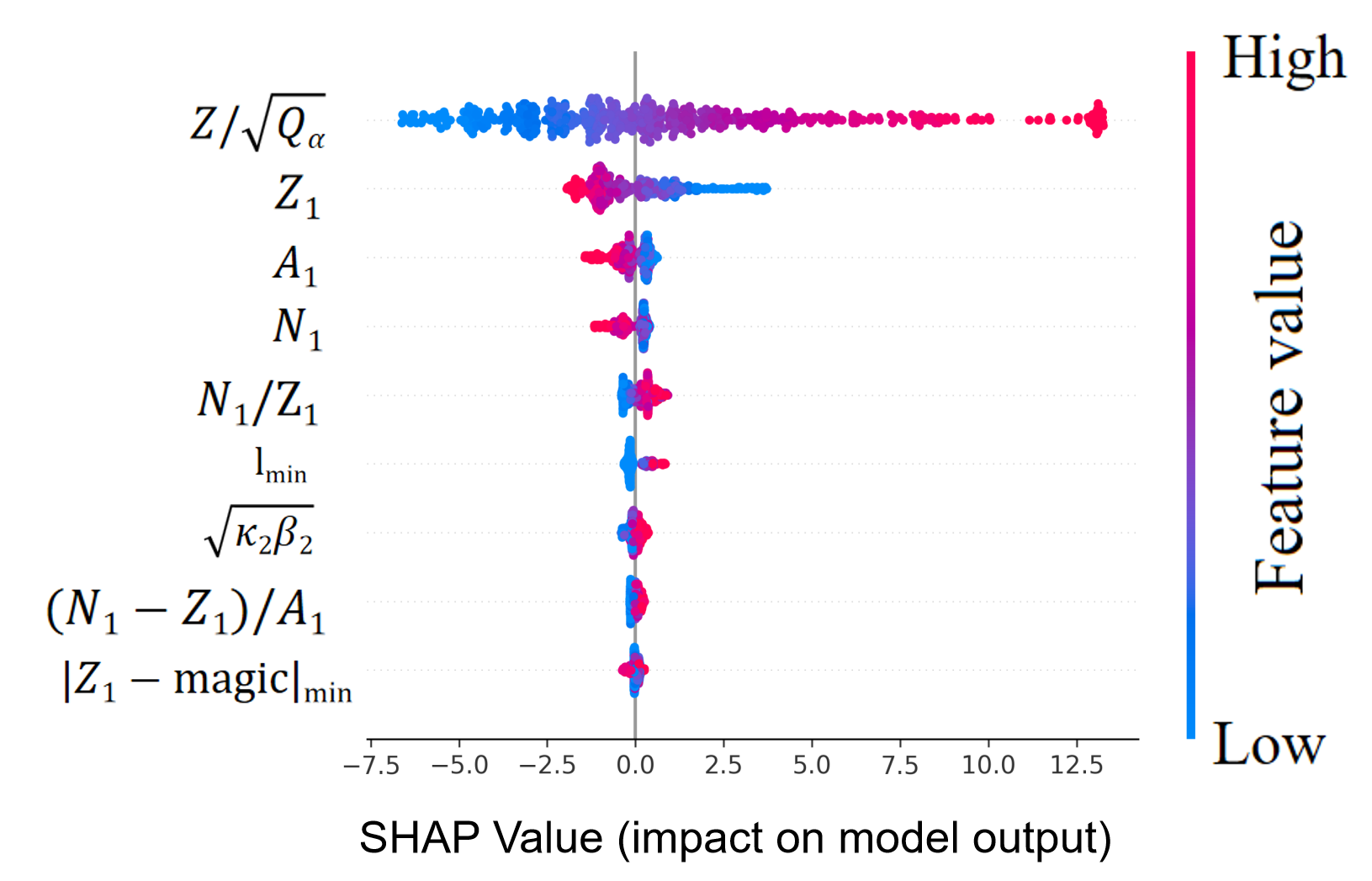}
\caption{SHAP-based feature importance for the $\alpha$-decay half-life 
prediction model.}
\label{fig:hl_shap}
\end{figure}

Since $\alpha$-decay half-lives are governed by multiple physical mechanisms—including decay energy, centrifugal-barrier effects, shell structure, nuclear deformation, and isospin asymmetry—it is essential to examine how these physical factors are utilized within the machine-learning model. To this end, we employ the SHAP (SHapley Additive exPlanations) interpretability framework to analyze the trained XGBoost model. SHAP enables a quantitative decomposition of the prediction into contributions from individual nuclear-structure features and provides insight into whether the learned dependencies are consistent with established nuclear-structure and decay systematics. The resulting feature-importance distribution is shown in Figure~\ref{fig:hl_shap}.

Figure~\ref{fig:hl_shap} presents the SHAP feature-importance distribution for the $\alpha$-decay half-life regression model. A clear hierarchical structure of physical contributions can be identified from the SHAP distribution.

Among all variables, the composite feature $Z/\sqrt{Q_\alpha}$ exhibits by far the largest SHAP amplitude, indicating that it provides the dominant contribution to the predicted $\log_{10}(T_{1/2})$. In the SHAP distribution, large values of $Z/\sqrt{Q_\alpha}$ (red points) are predominantly located on the positive side of the SHAP axis, implying an increase in the predicted half-life, whereas smaller values (blue points) tend to produce negative contributions. This behavior is fully consistent with the barrier-penetration picture of $\alpha$ decay and directly reflects the Geiger--Nuttall systematics, in which the decay probability is governed by the combined effect of the Coulomb barrier and the released decay energy. The dominance of $Z/\sqrt{Q_\alpha}$ therefore confirms that the model correctly captures the fundamental energy dependence of $\alpha$-decay half-lives.

Secondary contributions arise from nucleon-number related variables such as $Z_1$, $A_1$, and $N_1$. Although their SHAP amplitudes are smaller than that of the dominant feature, their distributions exhibit systematic patterns. These variables encode the global nuclear scale and nucleon configuration, which influence the Coulomb interaction, nuclear radius, and shell structure. Their non-negligible SHAP contributions indicate that the model incorporates nucleon-number systematics to refine the half-life prediction across different mass regions of the nuclear chart.

The minimum orbital angular momentum $\ell_{\min}$ provides a particularly transparent physical interpretation within the model. 
In $\alpha$ decay, transitions with $\ell_{\min}=0$ correspond to favored decays, whereas transitions with $\ell_{\min}>0$ are classified as unfavored decays because they require a non-zero orbital angular momentum transfer. 
In the SHAP distribution, larger values of $\ell_{\min}$ generally give rise to positive SHAP contributions, corresponding to an increase in the predicted $\log_{10}(T_{1/2})$, while $\ell_{\min}=0$ is associated with comparatively smaller or even negative contributions. 
This trend reflects the well-known centrifugal hindrance mechanism: when $\ell_{\min}>0$, an additional centrifugal barrier is introduced into the effective potential, which reduces the barrier-penetration probability and therefore prolongs the half-life. 
The SHAP behavior of $\ell_{\min}$ thus shows that the model is able to distinguish favored and unfavored decays in a physically meaningful way and correctly capture the role of angular-momentum hindrance in $\alpha$-decay systematics.

The deformation-related term $\sqrt{\kappa_2\beta_2}$ also produces noticeable contributions. Variations in this feature modify the effective shape of the Coulomb barrier and consequently affect the tunneling probability of the emitted $\alpha$ particle. The presence of non-zero SHAP amplitudes indicates that the model incorporates deformation effects to improve predictions in regions where nuclear shapes deviate from spherical symmetry.

Additional variables, including the isospin-asymmetry indicator $N_1/Z_1$, the relative neutron excess $(N_1-Z_1)/A_1$, and the distance to the nearest proton magic number, contribute at a smaller yet non-negligible level. These quantities reflect isospin asymmetry and shell-structure effects, which influence nuclear stability and cluster preformation. Although their individual impacts are weaker, they provide useful structural corrections that improve the predictive accuracy in specific regions of the nuclear chart, particularly near shell closures.

Overall, the SHAP analysis demonstrates that the XGBoost regression model does not behave as a purely numerical black-box interpolator. Instead, it reproduces the established hierarchical dependence of $\alpha$-decay half-lives on decay energy, nucleon-number systematics, angular-momentum hindrance, nuclear deformation, and shell structure. This physically consistent learning behavior explains the systematic improvement of the machine-learning model over traditional empirical formulas while remaining fully compatible with known nuclear-decay systematics.

\section{Acknowledgements}

This work is supported by Yunnan Provincial Science Foundation Project (No. 202501AT070067), Yunnan Provincial Xing Dian Talent Support Program (Young Talents Special Program, (Young Talents Special Program, No. XDYC-QNRC-2023-0162), Kunming University Talent Introduction Research Project (No. YJL24019), Yunnan Provincial Department of Education Scientific Research Fund Project (No. 2025Y1055 and 2025Y1042), the Special Basic Cooperative Research Programs of Yunnan ProvincialUndergraduate Universities’ Association (NO. 202101BA070001-144), the Program for Frontier Research Team of Kunming University 2023, National Natural Science Foundation of China (No. 12063006), National College Student Innovation and Entrepreneurship Training Program (No. 202511393012, 202511393013, and 202511393016), Yunnan Province College Student Innovation and Entrepreneurship Training Program (No. S202511393003, S202511393043, and S202511393044), and Xing Dingyu Academician Workstation of Yunnan Province (No. 202605AF350035).

\bibliographystyle{IEEEbib}
\bibliography{bib}
\end{document}